\documentclass[aps,prl,reprint,twocolumn,groupedaddress,showpacs]{revtex4-1}
\usepackage[utf8]{inputenc}
\usepackage[T1]{fontenc}

\usepackage{hyperref}
\usepackage{graphicx}
\usepackage{amsfonts}
\usepackage{amssymb}
\usepackage{amsmath}
\usepackage{mathrsfs}
\usepackage{commath}

\usepackage{txfonts}
\usepackage{lipsum}
\usepackage{color}
\usepackage{wasysym}

\usepackage{gensymb}
\usepackage{tabularx,ragged2e,booktabs,caption}
\usepackage{subfigure}
\usepackage{tabularx,capt-of}
\usepackage{multirow}

\usepackage{makecell,booktabs}

\definecolor{dark-green}{RGB}{0, 128, 0}
\newcommand{\piR}[1]{\textcolor{black}{#1}}

\usepackage{array}
\usepackage{braket}

\newcommand{\llo}{\boldsymbol{\lambda}_{0}}
\newcommand{\la}{\lambda}
\newcommand{\lla}{\boldsymbol{\lambda}}
\newcommand{\RR}{\boldsymbol{\mathcal{R}}}

\newcommand{\ii}{{\rm i}}
\newcommand{\ee}{{\rm e}}

\newcommand{\dd}{{\rm d}}

\usepackage{graphicx}

\begin{document}
\title{Symmetry-protected multifold exceptional points and their topological characterization}

\author{Pierre Delplace}
\affiliation{Univ Lyon, Ens de Lyon, Univ Claude Bernard, CNRS, Laboratoire de Physique, F-69342 Lyon, France}
\author{Tsuneya Yoshida}
\affiliation{Department of Physics, University of Tsukuba, Tsukuba, 305-8571 Japan}
\author{Yasuhiro Hatsugai}
\affiliation{Department of Physics, University of Tsukuba, Tsukuba, 305-8571 Japan}


\begin{abstract}
We investigate the occurrence of $n$-fold exceptional points (EPs) in non-Hermitian systems, and show that they are stable in $n-1$ dimensions in the presence of anti-unitary symmetries that are local in parameter space, such as e.g. parity-time (PT) or charge-conjugation parity (CP) symmetries. This implies in particular that 3-fold and 4-fold symmetry-protected EPs are stable respectively in 2 and 3 dimensions. The stability of such \textit{multofold} exceptional points (i.e. beyond the usual 2-fold EPs) is expressed in terms of the homotopy properties of a \textit{resultant vector} that we introduce. Our framework also allows us to rephrase the previously proposed $\mathbb{Z}_2$ index of PT and CP symmetric gapped phases beyond the realm of two-band models. We apply this general formalism to a frictional shallow water model that is found to exhibit  3-fold exceptional points associated with topological numbers $\pm1$. For this model, we also show different non-Hermitian topological transitions associated with these exceptional points, such as their merging and a transition to a regime where propagation is forbidden, but can counter-intuitively be recovered when friction is increased furthermore.
\end{abstract}
\maketitle

Since the discovery of graphene, Dirac points have been revealed to be an ubiquitous property of various two dimensional (2D) materials. The existence of such 2-fold degeneracies is enforced by symmetries \cite{Miert16,YoungKane15}, and their stability can be expressed with  winding numbers of the phase of the wave functions \cite{Ando98, Mikitik08, Fuchs10}. Similarly, non-hermitian Hamiltonians may exhibit degeneracy points of their complex eigenvalues, called exceptional points (EPs). The quest for EPs, their topological properties and  their physical implications, stimulated tremendous efforts in the past few years~\cite{Bergholtz19,Ashida_nHReview_arXiv20,Shen_EPs_PRL18,Zhou_EP-PHC_Science18,Kozii_EPs_arXiv17,Yoshida_EP_PRB18,Rausch_EPin1D_NJP20,Yoshida_EPs_PTEP20,Hofmann_EPs-ele_cir_PRR20, Shi16}. At an EP, the number of degenerated eigenvalues (the algebraic multiplicity $\mu$) is generically larger than the number of eigenvectors (the geometric multiplicity) leaving the Hamiltonian non-diagonalizable. Quite remarkably, 2-fold EPs (EP2s) do not require any symmetry to be stable in 2D, and this stability can be expressed in terms of a winding number~\cite{Shen_EPs_PRL18} associated to their complex eigenvalues \cite{Yang19}. This is however not the case for multi-fold exceptional points EP$\mu$s. Those were indeed shown to be unstable in 2D \cite{Yang19}, although stable in higher $D=2\mu-2$ dimension  \cite{Holler20, Bergholtz19}, so that EP$\mu$s beyond $\mu=2$ have remained overlooked. Still, a few examples of multifold EPs were surprisingly reported recently \cite{Xiao20,Hatano20,Haack21}, including in PT-symmetric systems \cite{ZhangPRA20}. This asks the crucial question of the conditions of existence of multifold EPs, in particular in dimensions $D\leq 3$, their robustness and the role of symmetries.

Here we answer this question by showing that  $\mu$-fold EPs have a \textit{codimension} $\mu-1$ provided certain anti-unitary symmetries are satisfied, meaning that they appear as isolated points in $ \mu-1$ dimensions.  The relevant symmetries consist in PT symmetry, CP symmetry and their generalizations, called pseudo-Hermiticity (psH) and pseudo-chirality (psCh), that  classify non-Hermitian EPs \cite{Kawabata19,Yoshida_EPs_PTEP20}. It follows that 3-fold and 4-fold EPs are stable in 2D and 3D respectively when such symmetries apply. Then, we propose a topological characterization for symmetry-protected EP$\mu$s through the homotopy properties of a \textit{resultant vector} that we introduce. Finally, we illustrate our theory by discussing a frictional fluid model that exhibits several EP3s with opposite windings. These EPs constitute a threshold beyond which the propagation of all the eigenmodes vanishes in a certain range of wavelength, but can counter-intuitively be recovered when increasing friction furthermore.

\piR{
To investigate $n$-fold EPs, we  focus on  $n \times n$ matrices  without loss of  generality, because $ n$-fold degeneracies are described by a $n \times n$ effective Hamiltonians $H(\lla)$, with  $\lla\in \mathbb{R}^D$. The degeneracies of the eigenvalues $E(\lla)$ are generically EPs, since arbitrary non-Hermitian matrices require fine tuning to be diagonalizable at the degeneracy point  \cite{SM}. Therefore, looking for $n$-fold EPs essentially amounts to searching for the conditions such that the characteristic polynomial $P_{\lla}(E)\equiv \det(H(\lla)-E)=a_n(\lla) E^n + a_{n-1}(\lla) E^{n-1} + \dots +a_1(\lla) E +a_0(\lla) $ has $n$-fold multiple roots and $n=\mu$. }

At an EP$\mu$ of energy $E_0$, $P_{\lla}(E)$ and its $\mu-1$ successive derivatives $P_{\lla}^{(j)}(E)\equiv \partial^{j} P_{\lla} / \partial E^j$ must vanish at $E=E_0$. This property is encoded in the zeros of the so-called \textit{resultant} $R_{P,P^{(j)}}(\lla)$. The resultant of two polynomials $P_1(E)$ and $P_2(E)$ is an elementary concept in algebra \cite{resultant}. It reads $R_{P_1,P_2}\equiv\det S_{P_1,P_2}$ where $S_{P_1,P_2}$ is the Sylvester matrix of $P_1$ and $P_2$. Importantly, this quantity vanishes if and only if the two polynomials $P_1$ and $P_2$ have a common root. Actually, we demonstrate (see \cite{SM}) that a polynomial $P(E)$ of degree $\mu$ has an $\mu$-fold multiple root if and only if the resultants of its successive derivatives $R_{P^{(j-1)},P^{(j)}}$ vanish. Put formally, \begin{align}
\label{eq:EP}
\begin{array}{c}
 \mu\times\mu\, \text{non-Hermitian matrices}\\ H(\lla)\, \text{have an EP}\mu \,
 \text{at}\ \lla=\llo
 \end{array}
 \Leftrightarrow
\begin{array}{c}
R_{P^{(j-1)},P^{(j)}}(\llo)=0   \\
\text{for}\  j=1\dots \mu-1 .
\end{array}
\end{align}

$R_{P^{(j-1)},P^{(j)}}(\lla)$ is in general a complex-valued function of $\lla$. The equivalence \eqref{eq:EP} thus yields $2(\mu-1)$ constraints to be satisfied, which gives the codimension of an EP$\mu$ in the absence of symmetry, in agreement with \cite{Holler20, Bergholtz19}. Note that the usual EP2s are  defined by $R_{P,P'}(\llo)=0$, where $R_{P,P'}(\lla)$ is proportional to the discriminant $\Delta$ of the characteristic polynomial. Since $\Delta$ is in general a complex function of $\lla$, the stability of EP2s  can be expressed by the winding of $\arg(R_{P,P'}(\lla))$ along a close circuit around $\llo$ in a 2D parameter space \cite{Yang19}. 

As we discuss now,  anti-unitary symmetries that are local in parameter space, have  important consequences. First, they decrease the codimension of the EPs. Second, they make the discriminant $\Delta$ real. Thus sgn($\Delta$) becomes a well-defined quantity that one can use to characterize a spontaneous symmetry breaking. The winding of $\arg(\Delta)$ becomes ill-defined, but we shall see that another natural homotopy property can be assigned to multifold EPs.

The symmetries we consider are 
\begin{subequations}
 \label{eq:sym}
\begin{align}
\label{eq:PT}    \text{PT-symmetry}\qquad        &U_{PT}H(\lla)U^{-1}_{PT}=H^*(\lla) \\
\label{eq:psH}   \text{pseudo-Hermiticity}\qquad  &U_{psH}H(\lla)U^{-1}_{psH}=H^\dagger(\lla)\\
\label{eq:CP}    \text{CP-symmetry}\qquad       &U_{CP}H(\lla)U^{-1}_{CP}=-H^*(\lla) \\
\label{eq:psC}   \text{pseudo-chirality}\qquad     &U_{psCh}H(\lla)U^{-1}_{psCh}=-H^\dagger(\lla)
     \end{align}
   \end{subequations}
where  $^*$ stands for complex conjugation, $^\dagger$ stands for Hermitian conjugation and the $U$'s are unitary operators. 
The consequences of these symmetries on the existence of EPs are readily obtained from the characteristic polynomial that must fulfill 
\begin{equation}
P_{\lla}(E)=\det(UH(\lla)U^{-1}-E) \ .
\label{eq:charac}
\end{equation}

Let us first proceed with PT-symmetry \eqref{eq:PT}. 
Then, the equation \eqref{eq:charac} implies $P_{\lla}(E)=a_n^*(\lla)E^n + a_{n-1}^*(\lla)E^{n-1} + \dots +a_1^*(\lla)E +a_0(\lla)^*$, so that $a_l^*(\lla)=a_l(\lla)$. All the coefficients $a_l(\lla)$ must therefore be real. 
Thus, the resultants $R_{P^{(j-1)},P^{(j)}}(\lla)$ are real too. This is due to the fact that the elements  of the Sylvester matrix $S_{P^{(j-1)},P^{(j)}}$ are essentially the coefficients $a_l$ (multiplied adequately by numbers of the form $n (n-1)\cdots(n-j+1)$ due to the $j$ successive derivatives) \cite{resultant}.
 The reality of the resultants implies that the number of conditions in \eqref{eq:EP} to have an EP$\mu$ reduces to $\mu-1$. By definition, the codimension of a PT-symmetry protected EP$\mu$ is therefore $\mu-1$. 

Importantly, this codimension remains $\mu-1$ for any of the symmetries introduced in \eqref{eq:sym}. This is obvious for pseudo-Hermiticity \eqref{eq:psH}, since taking the transpose of $H$ leaves invariant the determinant in Eq. \eqref{eq:charac}.
The case of CP-symmetry can be mapped onto the PT-symmetric one by the transformation $H\rightarrow -\ii H$, that  does not change the codimension of the degeneracies. Finally the pseudo-chiral case is deduced from the CP-symmetric one by invariance of Eq. \eqref{eq:charac} by taking the transpose of $H$.
One can finally rephrase this result as :
 $\mu$-fold complex degeneracies of a $\mu\times\mu$ Hamiltonian satisfying one of the local anti-unitary symmetries \eqref{eq:sym} appear as ''defects'' of dimension $d$ in a $D$-dimensional $\lla$-parameter space such that
\begin{align}
\label{eq:codim}
\text{codim(EP}\mu) \equiv  D-d  = \mu-1   \ .
\end{align}

Moreover, for each symmetry \eqref{eq:sym}, the resultants $R_{P^{(j-1)},P^{(j)}}$ are real  \cite{SM}, consistently with the $\mu-1$ constraints \eqref{eq:EP} to be satisfied. The reality of the resultants naturally generates  a  \textit{resultant vector}  $\tilde{\RR} $ of components $R_{P^{(j-1)},P^{(j)}}$, that maps the  $\lla$-space of parameters to $ \mathbb{R}^{\mu-1}$. When the dimension of the $\lla$-space is $\mu-1$, the homotopy properties of the map $\lla\rightarrow \tilde{\RR}/|\tilde{\RR}| $ can be used to characterize the EP$\mu$. 

In fact, a similar construction can be done from different resultants, so that another resultant vector, $\RR$, can be introduced. Indeed, the existence of an EP$\mu$ imposes resultants between other derivatives of $P(E)$ to vanish as well, and in particular 
\begin{align}
R_{P,P^{(j)}}(\llo)=0  \quad  \text{with} \ j=1\dots \mu-1 \ .
\end{align}
Actually, such a constraint, to be compared with \eqref{eq:EP}, is also a necessary and sufficient condition for an EP$\mu$ to exist, at least for $\mu=2,3$ and $4$ \cite{SM}. It turns out that those resultants are either real or purely imaginary. It is thus natural to define the resultant vector $\RR$ from its components 
\begin{align}
   \mathcal{R}_j\equiv \left\{
  \begin{array}{rl}
    \ R_{P,P^{(j)}}                     &  \text{-- PT and psH} \\
    \ (-\ii)^{n(n-j)} R_{P,P^{(j)}}   &  \text{-- CP and psCh} 
  \end{array}\right.
\end{align}
that depend on the symmetry, and where the coefficient $(-\ii)^{n(n-j)}$ guarantees the reality of $ \mathcal{R}_j$.

The two resultant vectors we have introduced have the same first component $\mathcal{R}_1$, which is proportional to the discriminant $\Delta$ of the characteristic polynomial $P_{\lla}(E)$ of degree $n$ as $R_{P,P'}(\lla)=(-1)^{n(n-1)/2}a_n(\lla)\, \Delta(\lla)$. The vanishing of the discriminant at a given $\llo$ indicates the existence of \textit{at least} two roots of $P_{\llo}(E)$. In the presence of a symmetry \eqref{eq:sym}, the discriminant is also a real quantity, since both $R_{P,P'}$ and $a_n$ are always real. Its sign (or equivalently that of $\mathcal{R}_1$) is therefore well defined and turns out to encode crucial properties about the complex eigenenergies $E_j$. For instance, it is well known that for a polynomial of degree $n=2$, $\Delta>0$ comes along with distinct real roots, while $\Delta<0$ indicates a pair of complex conjugated roots. These two behaviours are separated by a critical point $\Delta=0$ where the gap separating complex energy bands closes. More generally, a change in the number of complex conjugated roots is a property of the sign of the discriminant that generalizes beyond $n=2$ \cite{SM}. One can thus use sign($\Delta$) as a $\mathbb{Z}_2$ index to distinguish two different regimes for non-Hermitian Hamiltonians of arbitrary size that satisfy one of the symmetries \eqref{eq:sym}. The cases $n=2$ and $n=3$ are presented in table \ref{table:t}. Such an index encompasses the previously proposed topological invariants given by sgn$(\det H)$ [resp. sgn$( \det \ii H)$] for $2\times2$ PT [resp. CP] symmetric Hamiltonians~\cite{Gong_nHclass_PRX18,Okugawa_SPERs_PRB19,Yoshida_SPERmech_PRB19}. 

\begin{table}[h!]
\centering
\begin{tabular}{c|c|c|c}
degree $n$& sgn $\Delta$         &  PT / psH & CP / psCh \\
 \hline \hline
\multirow{3}*{$2$} & $+1$  & $E_i\in\mathbb{R} $, \ $E_1 \neq E_2$    & $E_1=-E_2^*$    \\
                              &  $0$  &   $E_1=E_2$                             &  $E_1=E_2$       \\
                              &$-1$  &  $E_1=E_2^*$    & $E_i\in \ii\mathbb{R}, \ $ $E_1 \neq E_2 $ \\ 
\hline
\hline
\multirow{3}*{$3$} &$+1$  & $E_i \in \mathbb{R} $, \ $E_1 \neq E_2 \neq E_3$    & $E_1=-E_2^*$  \ \text{and} \ $E_3\in \ii\, \mathbb{R}$   \\
                              & $0$  &   $E_1=E_2$   \ \text{and} \ $E_3\in\mathbb{R}$    &    $E_1=E_2$   \ \text{and} \  $E_3\in \ii\,\mathbb{R}$     \\
                              &$-1$  &  $E_1=E_2^*$  \ \text{and} \ $E_3\in \mathbb{R}$   & $E_i \in \ii \mathbb{R} $, \ $E_1 \neq E_2 \neq E_3$  
\end{tabular}
\caption{Properties of the eigenvalues $E_i$ of  $2\times2$ and $3\times3$ non-Hermitian matrix  as a function of the sgn$\Delta$ of its characteristic polynomial $\det(H-E)$ of degree $n$, in the presence of symmetries \eqref{eq:sym}.}
\label{table:t}
\end{table}

It is worth pointing out that a change of sign($\Delta$) reveals that the symmetry (\ref{eq:sym}) under consideration is \textit{spontaneously} broken. Let us recall this notion in PT-symmetric systems  \cite{Bender98}. Consider $H\ket{\psi}=E\ket{\psi}$, then $U_{PT}^{-1}\kappa \ket{\psi}$ is also an eigenstate of $H$ with eigenvalue $E^*$, where $\kappa$ is the complex conjugation operator. If  $U_{PT}^{-1}\kappa \ket{\psi}$ is proportional to $\ket{\psi}$, then $\ket{\psi}$ is an eigenstate of the $PT$ operator, and the  eigenenergies are real. In the other case, $\ket{\psi}$ is not an eigenstate of  $U_{PT}^{-1}\kappa$ and the eigenenergies come by pairs $(E,E^*)$. PT symmetry is then said to be \textit{spontaneously} broken : it still holds at the level of the Hamiltonian, but not at the level of the eigenstates.
The same reasoning applies to CP-symmetric and pseudo-chiral Hamiltonians, where eigenenergies come by either by pairs $(E,-E^*)$ or are purely imaginary. In any case, the spontaneous symmetry breaking is accompanied with a change of sign of the discriminant that governs the complex nature and the pairing of the eigenvalues.
%

Symmetry-protected multifold EPs are special points where the symmetry is spontaneously broken, since they also demand the vanishing of the other components $\mathcal{R}_j=0$. 
From a geometrical point of view~\cite{Yoshida_SPER_PRB19,Yoshida_SPERmech_PRB19,Kimura_SPES_PRB19,Yoshida_EPs_PTEP20}, the solutions $\lla$ of each equation $\mathcal{R}_j(\lla)=0$ define a ''manifold'' $\mathcal{M}_j$ of dimension $D-1$  in parameter space of dimension $D$ (i.e. codimension 1). Thus, the coordinates $\llo$ of EP$\mu$s define a space that consists in their mutual $\mu-1$ intersections.
For instance, in $D=2$ dimensions, $\mathcal{M}_1$ and $\mathcal{M}_2$ consist in curves in a plane, and their mutual intersections are points that correspond to EP3s. Importantly, such intersections are generically robust to perturbations. Let us illustrate this point on a concrete model, the rotating shallow water model  with friction.

%

The linearized rotating shallow water model describes waves propagating in a thin layer of fluid in rotation. It is currently used to describe atmospheric and oceanic waves over large distances where the Coriolis force, encoded into a parameter $f$, is relevant \cite{VallisBook}. We consider a non-Hermitian version of this model that reads $H \Psi = E \Psi$ with 
\begin{equation}
\label{eq:ham_sw}
H = 
    \begin{pmatrix}
    \ii \gamma_x  &   \ii f          &    k_x   \\
    -\ii f        &   \ii \gamma_y   &    k_y   \\
    k_x           &       k_y        &    \ii \gamma_N
    \end{pmatrix} \ ,
    \qquad 
    \Psi =
    \begin{pmatrix} 
    \delta u_x \\ \delta u_y \\ \delta h
      \end{pmatrix} 
\end{equation}
with $ \delta u_x$ and $\delta u_y$ the small variations of the horizontal fluid velocity, $\delta h$ a small variation of the fluid's thickness, $k_x$ and $k_y$ the in-plane wave numbers, $\gamma_x$ and $\gamma_y$  the Rayleigh friction terms, and $\gamma_N$ the Newtonian friction. 
The competition between the two friction terms plays for instance a crucial role in the phenomenon of super-rotation \cite{warneford_dellar_2017}. Actually, this minimal model equivalently describes active fluids where the thickness field $h$ is formally replaced by a pressure field. \cite{Souslov19}  In that context, it was shown that the $\gamma$ terms can also be negative, thus allowing for gain \cite{James}. We now show that the friction terms generate symmetry-protected EP3s.

\begin{figure}[h!]
\centering
\includegraphics[width=8.5cm]{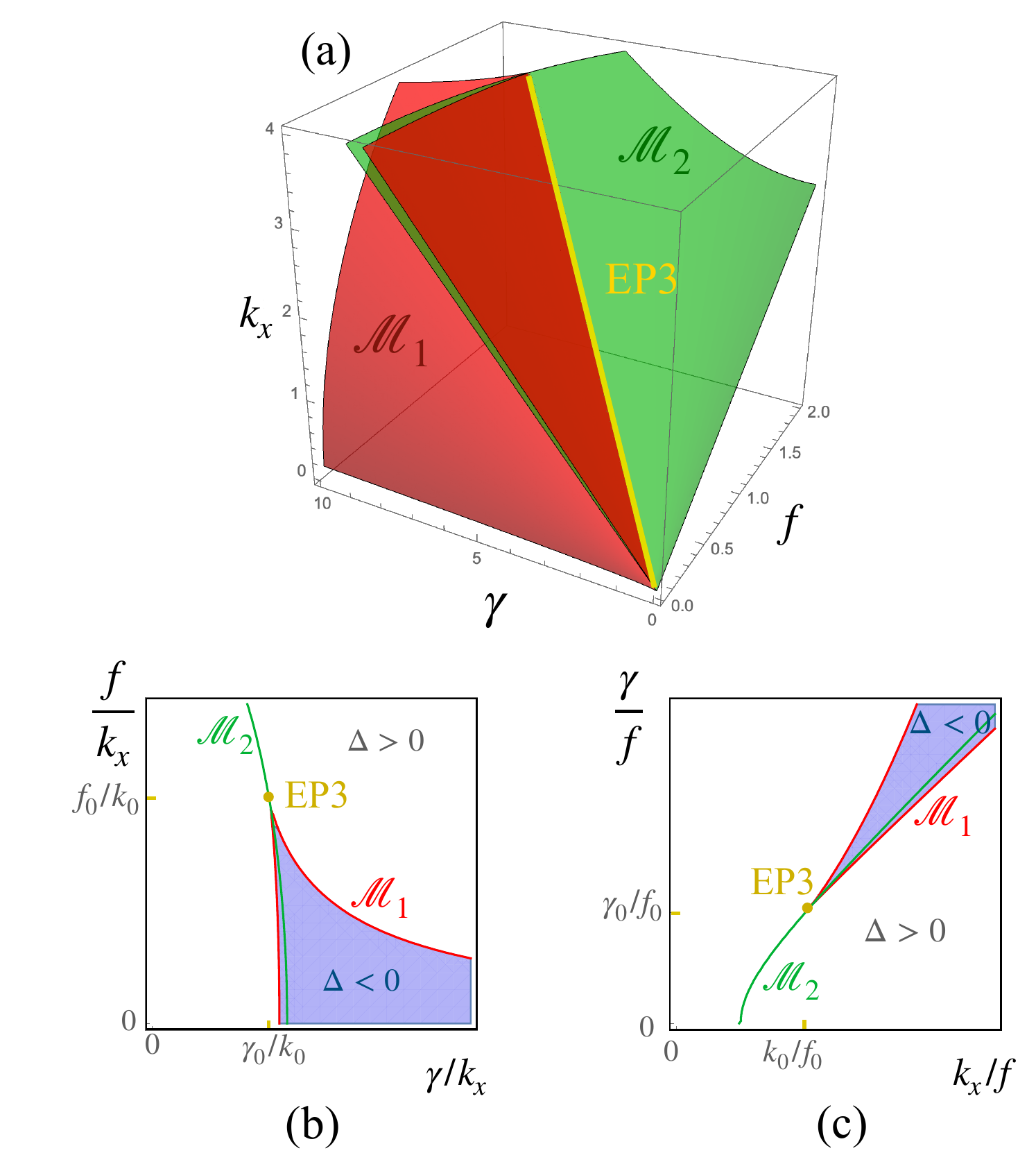}
\caption{ EP3s (yellow) of the frictional shallow water model in 3D (a) and 2D (b,c) parameter spaces. Those points result from the intersection of the surfaces $\mathcal{M}_1$ and $\mathcal{M}_2$ each of codimension 1.
}
\label{fig:EP3}
\end{figure}

Assuming isotropic friction $\gamma_x=\gamma_y\equiv \gamma_R$, the model remains rotational symmetric and can thus be simplified by fixing a direction (say $k_y=0$).
Moreover, any choice of $\gamma_R$ can be reabsorbed in $\gamma_N$ up to a global shift of the spectrum, that does not affect the existence of degeneracies, as $H=\gamma_R \mathbb{I} + \tilde{H}$ with
\begin{equation}
\label{eq:SW2}
\tilde{H}= 
    \begin{pmatrix}
    0  &   \ii f          &    k_x   \\
    -\ii f        &   0   &    0   \\
    k_x           &     0        &    \ii \gamma
    \end{pmatrix}
\end{equation}
where $\gamma \equiv \gamma_N -\gamma_R$. This matrix satisfies CP-symmetry \eqref{eq:CP}  with $U_{CP}=\text{diag}(1,1,-1)$, and one thus expects EP3s of codimension $2$, according to the relation \eqref{eq:codim}. Since the parameter space has a dimension $D=3$, with $\lla=\{k_x,f,\gamma\}$, the set of EP3s must consist in a manifold of codimension 2, that is a line.
For this specific model, the equation of this line can be derived explicitly from  $\mathcal{R}_1(\llo)=0$ and $\mathcal{R}_2(\llo)=0$, which yields $\gamma_{0}=\pm 3\sqrt{3}f_0$ and $k_0=\pm2\sqrt{2} f_0$.
Geometrically, this line of EP3s corresponds to the intersection of the two spaces $\mathcal{M}_1$ and $\mathcal{M}_2$, each of codimension 1.  $\mathcal{M}_1$ and $\mathcal{M}_2$ are thus surfaces in 3D (figure \ref{fig:EP3} (a)), and lines in  2D 
(Fig. \ref{fig:EP3} (b) and (c)). 

Remarkably, the space $\mathcal{M}_1 \cap \mathcal{M}_2$ of EP3s correspond to a fold of $\mathcal{M}_1$. Similar singular points to that shown in figure \ref{fig:EP3} (b) and (c) were found in \cite{Xiao20, Hatano20}. This suggests a quite fundamental property of EP3s, that can be addressed within catastrophe theory which classifies the possible kinks of curves and thus establishes their topological stability \cite{catastrophe_Castrigiano}.
Intuitively, such a singular shape can be traced back to the third order characteristic polynomial $P_{\lla}(E)=0$, since cubic curves are known as a basic example that displays such a \textit{catastrophe} when the three roots coincide. To characterize more precisely this singularity, we use the machinery of catastrophe theory and compute a list of invariants (called \textit{corank},  \textit{codimension} and \textit{determinacy}) that classify the catastrophe. Following \cite{catastrophe_Castrigiano,Chandrasekaran20}, we find a corank of $1$, a codimension of $1$ and a determinacy of $3$, which identifies EP3s as a \textit{fold} of $\mathcal{M}_1$. This is the simplest fundamental catastrophe in the classification of catastrophe theory.

\begin{figure}[h!]
\centering
\includegraphics[width=8.7cm]{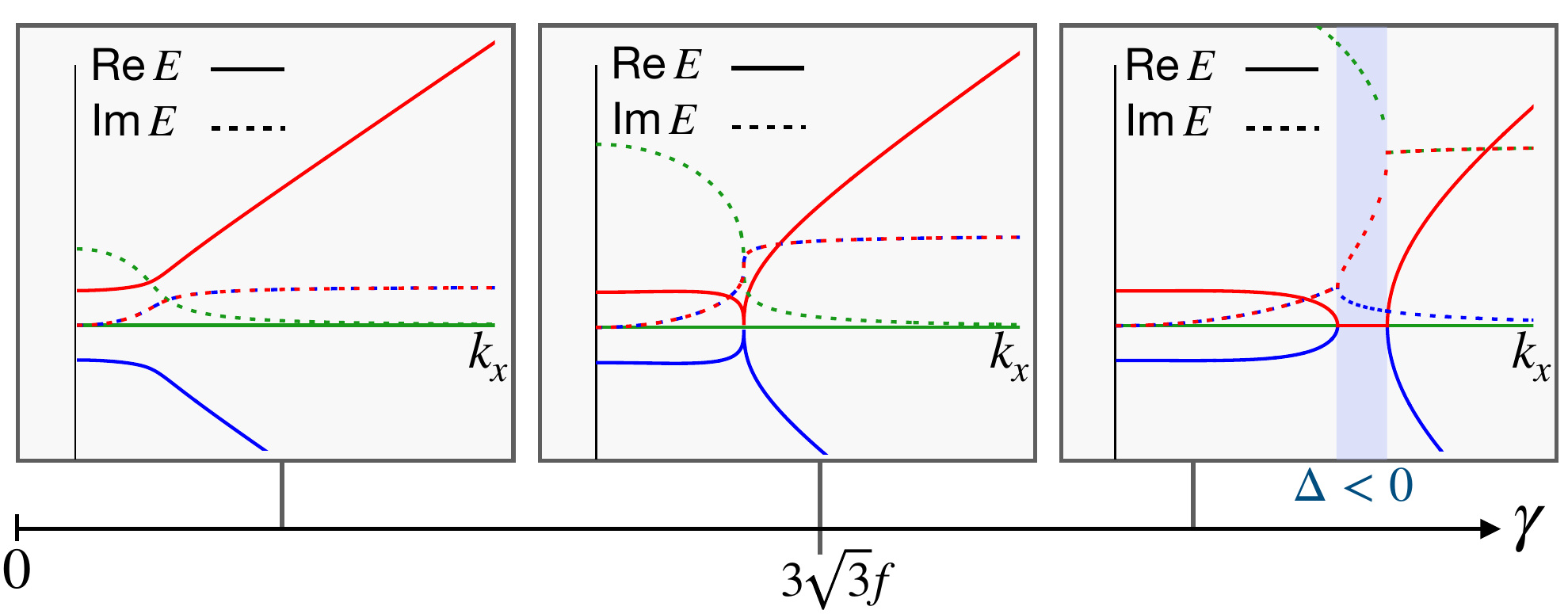}
\caption{Complex dispersion relation $E(k)$ of the shallow water model \eqref{eq:SW2} for different values of $\gamma$. An EP3 appears at the critical value $\gamma_0=3\sqrt{3}f$ for an arbitrary $f$ and its eigenfrequency is found to be $E_0=\ii \gamma_0/3$. Beyond this point, the discriminant $\Delta$ of the characteristic polynomial of $\tilde{H}$ becomes negative over a finite domain of wavenumbers, where the three eigenmodes are fully damped.}
\label{fig:dispersion}
\end{figure}

Before characterizing in more details the robustness of the EP3, let us first comment on the original physical behaviours revealed in figure \ref{fig:EP3}. 
Indeed, $\mathcal{M}_1$ indicates a change of sign of the discriminant, and thus, according to table \ref{table:t}, it denotes a transition between complex eigenvalues ($\Delta>0$) and purely imaginary eigenvalues $(\Delta<0)$, which, in our case, all have the same sign which is fixed by sgn$(\gamma)$. The eigenvalue spectrum of \eqref{eq:SW2} is shown in figure \ref{fig:dispersion}. It follows that the $\Delta<0$ domain corresponds to a regime where all the eigenmodes are fully evanescent (for $\gamma>0$) and thus where propagation is prohibited. Quite remarkably, while a first \textit{propagating} $\rightarrow$ \textit{non-propagating} transition happens when increasing friction, an even more striking \textit{non-propagating} $\rightarrow$ \textit{propagating} second transition occurs when increasing dissipation furthermore. In those phase diagrams, the EP3 appears as the threshold beyond which such transitions exist. 

%


One can characterize the stability of symmetry protected EPs furthermore by using the topological properties of the \textit{resultant vector} $\RR \in \mathbb{R}^{\mu-1}$ introduced above. For an EP3, it defines a map $\boldsymbol{\mathcal{R}}/|\boldsymbol{\mathcal{R}}| : \mathbb{R}^2 \backslash\{ \llo\}\rightarrow S^1$ whose homotopy properties are encoded into the integer-valued winding number
\begin{align}
\label{eq:W3}
W_3 =
\frac{1}{2\pi} \oint_{\mathcal{C}_{\lla}} \frac{1}{\norm{\RR}^2}\left(\mathcal{R}_1\frac{\partial \mathcal{R}_2}{\partial \la_\alpha} - \mathcal{R}_2\frac{\partial \mathcal{R}_1}{\partial \la_\alpha}\right)\dd \la_\alpha 
\end{align}
where the close circuit $\mathcal{C}_{\lla}$ surrounds the EP3 and where an implicit sum is taken over $\alpha=\{1,2\}$. 
It is worth noticing that, in lattice systems, $\mathcal{M}_1$ and $\mathcal{M}_2$ are necessarily closed loops in the 2D Brillouin zone, so that their intersections come by pairs with opposite $W_3$'s, which can be seen by a straightforward extension of the doubling theorem by Nielsen and Ninomiya  \cite{Nielsen_doubling1_NPhys1981,Nielsen_doubling2_NPhys1981} as recently generalized for non-Hermitian systems without symmetry \cite{Yang19}.

\begin{figure}[h!]
\centering
\includegraphics[width=9cm]{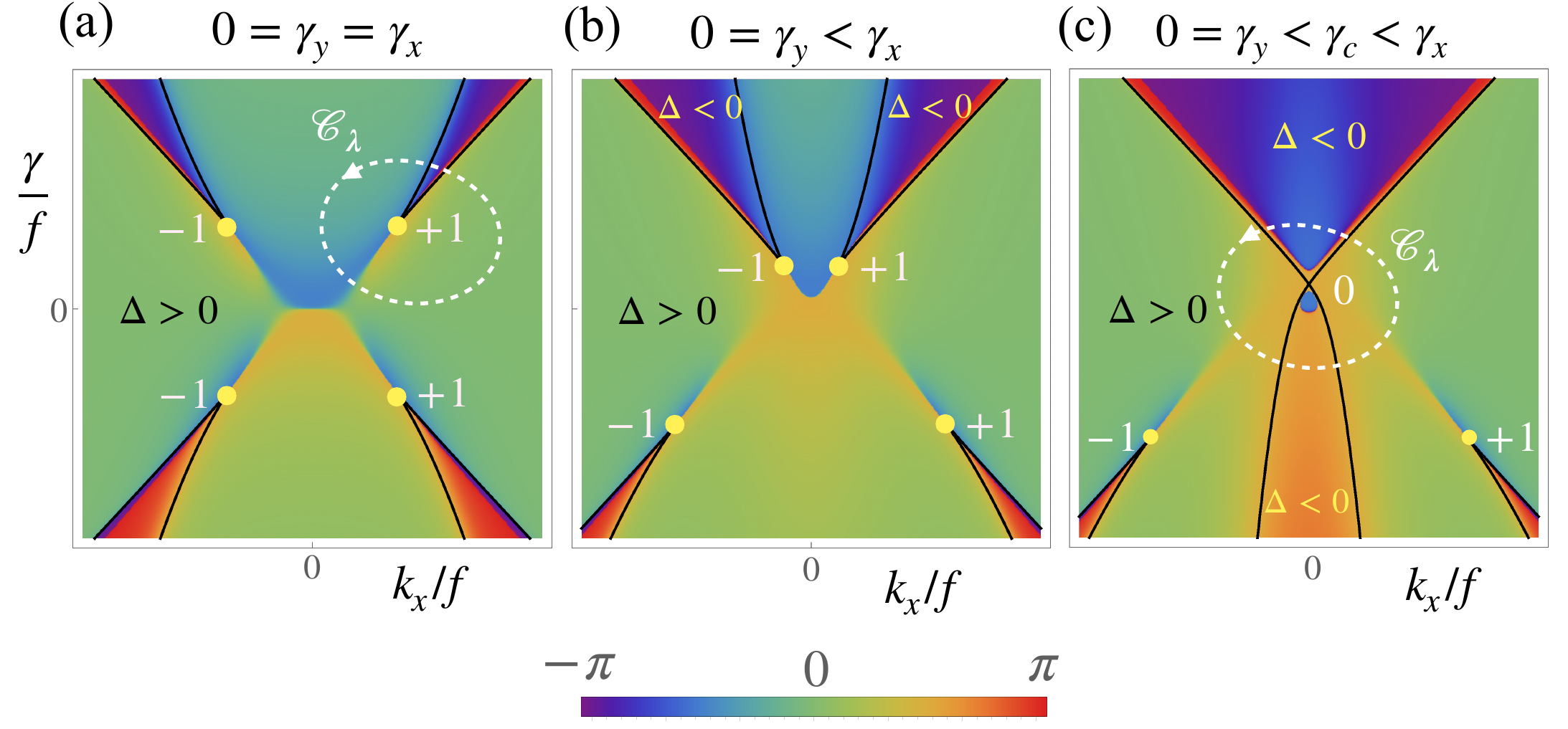}
\caption{Color plot of the relative angle between $\mathcal{R}_1$ and $\mathcal{R}_2$ in $(k_x/f,\gamma/f)$ parameter space for (a) no Rayleigh friction, (b) a small and (c) a large anisotropic Rayleigh friction. The winding of this angle along loops  surrounding the EP3's indicates 2 pairs of EP3s of charge $W_3=\pm 1$ until a critical value $\gamma_c$ of the anisotropy (c) beyond which a new domain with forbidden propagation ($\Delta<0$) emerges.}
\label{fig:EP}
\end{figure}

The invariant $W_3$ is the winding number of the relative angle between $\mathcal{R}_1$ and $\mathcal{R}_2$. For the frictional shallow water model, its value is shown in figure \ref{fig:EP} (a). It turns out that in this model, EP3s come by pairs $ W_3=\pm1$ at $k_x=\pm k_0$. This indicates that such pairs may possibly merge and annihilate each other if one moves them, similarly to Dirac and Weyl points in semimetals \cite{DieltPRL08,Wunsch_2008,MontambauxPRB09,DelplaceEPL12}. Since the position of the EP3s is fixed in the isotropic case, a mechanism involving their motion necessarily breaks rotational symmetry. This can be achieved by considering now $\gamma_x \neq \gamma_y$, as shown in figure \ref{fig:EP} (a) where one pair  ($\gamma_N/f>0$) get closer while the other pair ($\gamma_N/f<0$) is pulled away. When increasing $\gamma_x$ furthermore, the points of the pair at $\gamma_N/f>0$ finally merge. Surprisingly enough,  a new domain of non-propagating waves ($\Delta<0$) emerges for $\gamma_N<0$ (see Fig. \ref{fig:EP} (b)).A similar analysis can be carried out from the other resultant vector $\tilde{\RR}$ introduced above, but its winding number is found to be zero for that model.

More generally, multifold ($\mu\geqslant3$) symmetry-protected EPs can be characterized by the homotopy properties of their resultant vectors $\RR$ (or $\tilde{\RR}$). Let us fix $d=0$, so that the EP is an isolated point $\llo \in \mathbb{R}^{\mu-1}$, according to \eqref{eq:codim}. Then the resultant vector defines the map $\boldsymbol{\mathcal{R}}/|\boldsymbol{\mathcal{R}}| : \mathbb{R}^{\mu-1} \backslash\{ \llo\} \rightarrow  S^{\mu-2}$ whose   degree\footnote{Not to be confused with the degree of the characteristic polynomial.} $W_\mu \equiv \deg \RR \in \mathbb{Z}$ is well-defined and reads \cite{DubrovinBook}
\begin{align}
   W_\mu = \sum_{\lla_i} \text{sgn} \left(\frac{\partial \mathcal{R}_j}{\partial \la_p}\right)\bigg|_{\RR(\lla_i)=\RR_0} 
\end{align}
where $\RR_0$ is an arbitrary point where the Jacobian matrix $(\partial \mathcal{R}_j/\partial \la_p)$ does not vanish. This homotopy invariant is a topological property of multifold symmetry-protected EPs. 
The expression \eqref{eq:W3} of the winding number $W_3$ for EP3s is the simplest example of such invariants. 

Our analysis opens various novel perspectives in the control of topological properties of non-Hermitian systems. In future works, it could be interesting to investigate symmetry-protected EP4s that should appear in three-dimensional parameter spaces, and whose topological charge $W_4$ takes the form of a wrapping number of the resultant vector
\begin{align}
W_4  = \frac{1}{4\pi} \oint_{\mathcal{S}_{\lla}} \frac{\RR}{\norm{\RR}^3}.\left(\frac{\partial \RR}{\partial \la_\alpha}\times\frac{\partial \RR}{\partial \la_\beta}   \right)\dd \la_\alpha \wedge \dd \la_\beta    
\end{align}
In particular, this suggests the existence of stable non-Hermitian versions of 3D Dirac points in sharp contrast with both Hermitian 3D Dirac points and non-Hermitian Weyl points that were both found to be unstable.

\begin{acknowledgments}-- \textbf{Acknowledgments} -- We are grateful to Antoine Venaille for many stimulating discussions about the shallow water model.
This work is supported by JSPS Grant-in-Aid for Scientific Research on Innovative Areas "DiscreteGeometric Analysis for Materials Design'': Grant No. JP20H04627.
This work is also supported by JSPS KAKENHI Grants No.~JP17H06138, and No.~JP19K21032.
\end{acknowledgments}

\bibliography{bibliography}

\appendix

\section{SM1 -- Accidental diagonalization}

By analyzing a $2\times 2$ matrix, we illustrate that fine tuning is necessary to make Hamiltonians diagonalizable.

Consider a one-dimensional $PT$-symmetric system whose non-Hermitian Hamiltonian is written as
\begin{eqnarray}
H(k)&=& 
\left(
\begin{array}{cc}
 1        &  k +\lambda \\
k-\lambda & 1
\end{array}
\right)
\end{eqnarray}
with $\lambda \geq 0$ and where $k$ denotes the momentum.
Diagonalizing this matrix yields the eigenvalues
\begin{eqnarray}
E_\pm(k) &=& 1\pm \sqrt{k^2-\lambda^2}
\end{eqnarray}
and the eigenvectors
\begin{align}
\psi_\pm = 
\begin{pmatrix}
k+\lambda \\
E_\pm(k) -1 
\end{pmatrix} \ .
\end{align}
Those eigenvectors are clearly linearly independent, except when the two bands touch, which occurs at $k= \lambda$. In that case, one has $\psi_+=\psi_-$, and the characteristic polynomial $P(E)=E^2-2E-k^2+\lambda^2+1$ has a common root $E=1$ with its derivative with respect to $E$, $P'(E)=2(E-1)$. At this degeneracy point, fine tuning of $\lambda$ is necessary in order to make $H(k)$ diagonalizable.
Indeed,  $H(k=\lambda)$ reads
\begin{eqnarray}
H(\lambda)&=& 
\left(
\begin{array}{cc}
 1        &  2\lambda \\
 0        &   1
\end{array}
\right)
\end{eqnarray}
which  has a single eigenvector, proportional to
\begin{eqnarray}
\left(
\begin{array}{c}
1  \\
0
\end{array}
\right)
\end{eqnarray}
when $\lambda>0$.
Thus, the dimension of the associated eigenspace (i.e. the geometric multiplicity) is reduced to $1$, while the algebraic multiplicity $\mu=2$. This two-fold degeneracy point is therefore an exceptional point. 
The diagonalisation of $H(k=\lambda)$ is then recovered if we tune
$\lambda=0$, and thus we simply have
\begin{eqnarray}
H(\lambda=0)&=& 
\left(
\begin{array}{cc}
 1        &   0 \\
 0        &   1
\end{array}
\right),
\end{eqnarray}
which is Hermitian, and which has two linearly independent eigenvectors
\begin{eqnarray}
\left(
\begin{array}{c}
1  \\
0
\end{array}
\right)
\quad
\text{and}
\quad
\left(
\begin{array}{c}
0  \\
1
\end{array}
\right) 
\end{eqnarray}
so that the algebraic and geometric multiplicities \textit{accidentally} coincide.

\section{SM2 -- Proof that $p$-fold exceptional points imply   $R_{P^{(j-1)},P^{(j)}}=0$ and  $R_{P,P^{(j)}}=0$ for $j=1\dots p-1$ .}

Consider two polynomials 
\begin{align}
&A(X) = a_n X^n + a_{n-1}X^{n-1} + \cdots + a_1 X+a_0\\
&B(X) = b_m X^m + b_{m-1}X^{m-1} + \cdots + b_1 X+b_0
\end{align}
of roots $\alpha_i$ and $\beta_i$ respectively, with $i=1\cdots n$ and $j=1\cdots m$, and where $a_n$ and $b_m$ are assumed to be non-zero.Then, let us recall that, by definition, their resultant reads
\begin{align}
R_{A,B} =  a_n^m b_m^n \prod_{i,j} (\alpha_i -\beta_j) \ .
\end{align}
It is clear from this expression that the resultant vanishes when the two polynomials $A$ and $B$ have \textit{at least} a common root.

Consider now a polynomial $P(X)$ of degree $n$
\begin{align}
P(X) = \prod_{i=1}^{n} (X-\alpha_i)
\end{align}
If this polynomial has a $p$-fold multiple-root $a=\alpha_1=\alpha_2 = \cdots = \alpha_p$, i.e.
\begin{align}
P(X) = (X-a)^p \prod_{i=p+1}^{n} (X-\alpha_i) 
\end{align}
then any of its  $j=1\cdots p-1$ derivative $P^{(j)}(X)$ factorizes by $X-a$, and has therefore $a$ has a root. Interpreting now $P$ has the characteristic polynomial of a non-Hermitian Hamiltonian $H$, it follows that 
\begin{align}
H\, \text{has a $p$-fold EP} \implies R_{P^{(q)},P^{(j)}} = 0 \quad \text{for}\, q,j=1\cdots p-1 \ .
\label{eq:implication1}
\end{align}

\section{SM3 -- Proof of equation (1) of the main text.}
Equation (1) of the main text is an equivalence. The ''if'' part of this equivalence follows from \eqref{eq:implication1}  by taking $q=j-1$. Here we demonstrate the ''only if'' part,
that is $R_{P^{(j-1)},P^{(j)}}=0$ for $j=1\dots n-1$ implies an EP$n$.
We know that this statement is true for $n=2$ (in that case the resultant is the discriminant).
We then proceed by induction. Let us assume that the implication 
\begin{align}
\begin{array}{c}
R_{P^{(j-1)},P^{(j)}}=0   \\ 
\text{for}\ j=1\dots n-1 
\end{array}
\implies 
\begin{array}{c}
 P\ \text{has a multiple} \\ 
 \text{root of order} \ n
 \end{array}
 \label{eq:implication2}
\end{align}
is true for polynomials $P$ of degree $n$, and let us assume that
\begin{align}
R_{P^{(j-1)},P^{(j)}}=0 \qquad \text{for} \ j=1\dots n 
\end{align}
for polynomials $P$ of degree $n+1$.
We then rewrite those equations as
\begin{align}
&R_{P,P'}=0 \\
&R_{(P')^{(j-1)},(P')^{(j)}}=0 \qquad \text{for}\, j=1\dots n-1 \ .
\end{align}
Since $P'$ is of degree $n$, the implication \eqref{eq:implication2} is true, and therefore
\begin{align}
P'=A(X-a)^n
\end{align}
where $A$ is a constant, so that 
\begin{align}
P=\frac{A}{n+1}(X-a)^{n+1} + B
\end{align}
with $B$ another constant.
Next, since $R_{P,P'}=0$, then $a$ must be a root of $P$, and thus $B=0$. It follows that $P$ as a multiple root  of order $n+1$, which completes the proof.
 
This, together with the result of the previous paragraph, demonstrates the equivalence
 (1) of the main text.

\section{SM4 -- Proof of the equivalence between the existence of an EP$\mu$ and equation (5) of the main text for $\mu=3$ and $\mu=4$}
Equation \eqref{eq:implication1} implies that if an EP$\mu$ exists, then $R_{P,P^{(j)}}=0$ with $j=1, \dots, \mu-1$ (by taking $q=0$).
We show now that the converse is also true (at least) for $3$-fold and $4$-fold EPs, as discussed in the main text.
We can restrict the discussion to cases where the order of the EP is that of the polynomial, that is, $\mu=n$. 

\subsection{Proof that $R_{P,P^{(j)}}=0$ with $j=1,2$ implies an EP$3$}
The characteristic polynomial reads (up to a global non-zero factor that we shall ignore)
\begin{align}
P(X) = (X-\alpha_1)(X-\alpha_2)(X-\alpha_3) 
\end{align}
and its derivatives are
\begin{align}
\begin{aligned}
P'(X) = &(X-\alpha_1)(X-\alpha_2) + (X-\alpha_2)(X-\alpha_3) +\\ & + (X-\alpha_3)(X-\alpha_1) 
\end{aligned}
\label{eq:p3'}\\
\begin{aligned}
P''(X) = &2[ (X-\alpha_1) + (X-\alpha_2) + (X-\alpha_3) ]  \ .
\end{aligned}
\label{eq:p3''}
\end{align}
We then assume that 
\begin{align}
\label{eq:R3pp'}
&R_{P,P'} = 0\\
\label{eq:R3pp''}
&R_{P,P''} = 0 \ .
\end{align}
Let us show that the three roots $\alpha_i$ are therefore equal. This can be performed with basic algebra.

First, \eqref{eq:R3pp'} implies that a root of $P'$, that we shall denote by $X_1$ satisfies $X_1=\alpha_1$ or $X_1=\alpha_2$ or $X_1=\alpha_3$. Let us fix fix $X_1=\alpha_1$ without loss of generality.
Inserting $X_1=\alpha_1$ into \eqref{eq:p3'} yields
\begin{align}
(\alpha_1-\alpha_2)(\alpha_1-\alpha_3) = 0
\label{eq:p3:1}
\end{align} 
which imposes $\alpha_1=\alpha_2$ or $\alpha_1=\alpha_3$. We shall consider those two possibilities, but first, let us translate the condition \eqref{eq:R3pp''}, which means that a root $X_2$ of $P''$ must be either $\alpha_1$ or $\alpha_2$ or $\alpha_3$.

Let us assume first that $X_2$ is a priori different than $X_1$, and take thus $X_2=\alpha_2$ without loss of generality. It then follows from \eqref{eq:p3''} that
\begin{align}
2\alpha_2 - \alpha_1 - \alpha_3 =0\ .
\end{align}
Then, if we assume from \eqref{eq:p3:1} that $\alpha_1=\alpha_2$, then it follows from the equation above that $\alpha_2=\alpha_3$. Thus $X_2$ and $X_1$ must be the same root, and $\alpha_1=\alpha_2=\alpha_3$.
If we assume instead  from \eqref{eq:p3:1} that $\alpha_1=\alpha_3$, then it follows again that $\alpha_2=\alpha_3$, and the same conclusion applies. Therefore $X_1$ cannot be different from  $X_2$.

Let us assume now that $X_2=X_1$, that is $X_2=\alpha_1$. In that case, it follows from \eqref{eq:p3''} that
\begin{align}
2\alpha_1 - \alpha_2 - \alpha_3 =0 \ .
\end{align}
Again, if we assume now from \eqref{eq:p3:1} that $\alpha_1=\alpha_2$,  this equation yields $\alpha_2=\alpha_3$. And if we assume instead  from \eqref{eq:p3:1} that $\alpha_1=\alpha_3$, one also get $\alpha_2=\alpha_3$. 

So, in any case, $\alpha_1=\alpha_2=\alpha_3$, that is, the polynomial $P$ has a triple root if \eqref{eq:R3pp'} and \eqref{eq:R3pp''} are satisfied. Since we shown previously that if $P$ has a triple root then $R_{P,P'} = R_{P,P''} = 0$, this proves the equivalence between these two statements.

\section{SM5 -- Proof that $R_{P,P^{(j)}}=0$ with $j=1,2,3$ implies an EP$4$}

The demonstration of the equivalence for $n=4$ follows the same line as that for $n=3$, but is a bit longer. We have
\begin{align}
P(X) &=  (X-\alpha_1)(X-\alpha_2)(X-\alpha_3) (X-\alpha_4)
\label{eq:p4}\\
P'(X) &= (X-\alpha_1)(X-\alpha_2)(X-\alpha_3) + (X-\alpha_1)(X-\alpha_2)(X-\alpha_4) \notag \\  + &(X-\alpha_1)(X-\alpha_3)(X-\alpha_4) + (X-\alpha_2)(X-\alpha_3)(X-\alpha_4)
\label{eq:p4'}\\
P^{(2)}(X) &= 2[(X-\alpha_1)(X-\alpha_2)+(X-\alpha_1)(X-\alpha_3) \notag \\ &+(X-\alpha_1)(X-\alpha_4)+(X-\alpha_2)(X-\alpha_3) \notag \\ &+(X-\alpha_2)(X-\alpha_4)+(X-\alpha_3)(X-\alpha_4) ]
\label{eq:p4''}\\
P^{(3)}(X)& = 6 (X-\alpha_1+ X-\alpha_2+ X-\alpha_3 + X-\alpha_4)
\label{eq:p4'''}
\end{align}
We then assume that 
\begin{align}
\label{eq:r4}
&R_{P,P'} = 0\\
&R_{P,P^{(2)}} = 0 \\
&R_{P,P^{(3)}} = 0 \ .
\end{align}
and we want to show that the  four roots $\alpha_i$ are necessarily equal. 
The equations \eqref{eq:r4} imply that each $P^{(j)}(X)$ shares a common root $X_j$  with $P(X)$, which are a priori different from each others. 
We can then choose $X_j = \alpha_j$ without loss of generality, which implies 
\begin{align}
&(\alpha_1-\alpha_2)(\alpha_1-\alpha_3)(\alpha_1-\alpha_4)=0 \label{eq:p4:1}\\
&(\alpha_2-\alpha_1)(\alpha_1-\alpha_3)+(\alpha_2-\alpha_1)(\alpha_2-\alpha_4) \notag \\& \qquad +(\alpha_2-\alpha_3)(\alpha_2-\alpha_4)=0 \label{eq:p4:2}\\
&3\alpha_3= \alpha_1 + \alpha_2 +\alpha_4 \label{eq:p4:3}
\end{align}

Equation \eqref{eq:p4:1} is satisfied when $\alpha_1=\alpha_2$, or $\alpha_1=\alpha_3$ or $\alpha_1=\alpha4$. Let us consider first the case where $\alpha_1=\alpha_2$. Then, according to \eqref{eq:p4:2}, it follows that 
\begin{align}
(\alpha_2-\alpha_3)(\alpha_2-\alpha_4)=0
\end{align}
which means that at least two of the roots $X_j$ must be equal.
We thus have two subcases to consider. If on the one hand $\alpha_2=\alpha_3$, then equation \eqref{eq:p4:3} implies $\alpha_3=\alpha_4$, and thus the four roots $\alpha_j$ are equal, because $\alpha_1=\alpha_2$. If  on the other hand  $\alpha_2=\alpha_4$, then equation \eqref{eq:p4:3} agains implies that $\alpha_3=\alpha_4$ and one reaches the same conclusion about the equalities between the roots.

From equation \eqref{eq:p4:1}, we could instead assume that $\alpha_1=\alpha_3$. In that case, equation \eqref{eq:p4:2} implies
\begin{align}
(\alpha_2-\alpha_1) (3\alpha_2-\alpha_1-2\alpha_4)  =0 \ .
\end{align}
One then has to examine to possibilities. If $\alpha_2=\alpha_1$, then equation \eqref{eq:p4:3} readily implies that $\alpha_3=\alpha_4$, and thus all the roots are equal. If we assume instead that $\alpha_1=3\alpha_2-2\alpha_4$, then  \eqref{eq:p4:3}  yields $\alpha_2=\alpha_4$, and the four roots are thus equal.
This demonstrates that if the three resultants in \eqref{eq:r4} vanish, then $P$ has a multiple root $\alpha_1=\alpha_2=\alpha_3=\alpha_4$ of multiplicity $4$.

As a conclusion, we have demonstrated the following equivalence 
\begin{align}
\begin{array}{c}
p\times p\, \text{non-Hermitian Hamiltonians}\\  \text{have a $p$-fold EP}
\end{array}
 \Leftrightarrow 
 \begin{array}{c}
R _{P,P^{(j)}} = 0\\ \text{for}\, j=1\cdots p-1 
\end{array}
\end{align}
when $p=3$ or $4$. (The case $p=2$ being the standard one.)

\section{SM6 -- Proof of the reality of the resultants $R_{P^{(j-1)},P^{(j)}}$ in the presence of a local anti-unitary symmetry }

When PT-symmetry (3a) or pseudo-Hermiticity (3b) is satisfied, the coefficients $a_l$ are real, and therefore the resultants are real as well.

In contrast, CP-symmetry (3c) implies $a_l=(-1)^{n+l}a_l^*$. This means that a polynomial of degree $n$, has coefficients $a_n\in \mathbb{R}$, $a_{n_1}\in \ii \mathbb{R}$, $a_{n_2}\in \mathbb{R}$ and so on. Let us then perform the change of variable $E\rightarrow -\ii E$, so that
\begin{align}
\begin{array}{rl}
P^{(j)}(-\ii X) \in \mathbb{R}                & \text{if} \quad n-j \,  \text{even} \\
-\ii P^{(j)}(-\ii X) \in \mathbb{R}           & \text{if} \quad n-j  \, \text{odd} \\
-\ii P^{(j-1)}(-\ii X) \in \mathbb{R}        & \text{if} \quad n-j \,  \text{even} \\
P^{(j-1)}(-\ii X) \in \mathbb{R}             & \text{if} \quad n-j  \, \text{odd} \\
\label{eq:machin}
\end{array}
\end{align}

 We now use the textbook formula  of the resultant between two polynomials $P_1$ and $P_2$ of degree $d_1$ and $d_2$ \cite{resultant}
\begin{align}
R_{\xi_1 P_1(\lambda X), \xi_2 P_2(\lambda X)}  = \xi_1^{d_2} \xi_2^{d_1} \lambda^{d_1 d_2}R_{P_1(X),P_2(X)}.
\label{eq:resultant_formula}
\end{align}
where $\lambda$, $\xi_1$ and $\xi_2$ are arbitrary numbers, to get
\begin{align}
 \label{eq:truc}
 R_{\xi_1P^{(j-1)}(-\ii E),\xi_2 P^{(j)}(-\ii E)} & =  (-\ii)^{(n-j+1)(n-j)}\\ \notag& \qquad \times \xi_1^{n-j} \xi_2^{n-j+1} R_{P^{(j-1)}(E),P^{(j)}(E)}  \ .
 \end{align}
Note that $(n-j+1)(n-j)$ is always even for any $n$ and $j$, so $(-\ii)^{(n-j+1)(n-j)}$ is real. 

Let us assume that $n-j$ is even. Then, the resultant $ R_{-\ii P^{(j-1)}(-\ii E), P^{(j)}(-\ii E)} $ is real because of \eqref{eq:machin}. Then, it follows from \eqref{eq:truc} for $\xi_1=-\ii$ and $\xi_2=1$ that $R_{P^{(j-1)}(E),P^{(j)}(E)}$ is real.

Let us assume now that $n-j$ is odd. In that case, one can consider the resultant $ R_{P^{(j-1)}(-\ii E), -\ii P^{(j)}(-\ii E)} $, that is real, because of \eqref{eq:machin}. We then apply the formula \eqref{eq:truc} for $\xi_1=1$ and $\xi_2=-\ii$ from which we deduce that $R_{P^{(j-1)}(E),P^{(j)}(E)}$ is real. 

We have demonstrated that the resultants $R_{P^{(j-1)}(E),P^{(j)}(E)}$ are always real when CP  symmetry (and thus pseudo-chirality) is satisfied. Those resultants are therefore always real when either of the four symmetries discussed in the main text is satisfied.

\section{SM7 -- Proof of Eq. 6 of the main text.}

To proof Eq. (6), let us substitute Eq. (3c) or (3d) in equation (4). It yields $a_l=(-1)^{n+l}a_l^*$.
The conditions for $R_{P,P^{(j)}}$ to be real thus depend on the parity of $n$ and $j$. To derive them, it is useful to introduce  the polynomial $\tilde{P}(E)\equiv P(-\ii E)$. This polynomial has real coefficients if $n$ is even and imaginary coefficients if $n$ is odd. Let us discuss each these two cases separately.
\begin{itemize}
    \item 
If $n$ is even, then
the resultants $R_{\tilde{P},\tilde{P}^{(j)}}$ are real for every $j$. They  are also related to the resultants of the original polynomial $P(E)$ through a formula that we recall : the resultant $R_{P_1(X),P_2(X)}$ of two polynomials $P_1(X)$ and $P_2(X)$ of degree $d_1$ and $d_2$, and that $R_{P_1(\lambda X),P_2(\lambda X)}$ of these same polynomials expressed in the variable $\lambda X$, are related by $R_{P_1(\lambda X),P_2(\lambda X)}=\lambda^{d_1 d_2} R_{P_1(X),P_2(X)}$. Applying this formula in our case simply yields
$R_{\tilde{P},\tilde{P}^{(j)}}=(-\ii)^{n(n-j)} R_{P,P^{(j)}}$. It follows that $R_{P,P^{(j)}} \in \mathbb{R}$ for $n$ even. 
\item If $n$ is odd, then the polynomial $\ii \tilde{P}$ has real coefficients, and its resultant is therefore real too. Due to the multiplication by $i$, one has now the relation $R_{\ii \tilde{P},\ii \tilde{P}^{(j)}}=\ii^{2n-j} (-\ii)^{n(n-j)} R_{P,P^{(j)}}$. Since $n$ is assumed to be odd, then $R_{P,P^{(j)}} \in \ii \mathbb{R}$ for $j$ even and $R_{P,P^{(j)}} \in \mathbb{R}$ for $j$ odd. 
\end{itemize}
Thus, to summarize, $R_{P,P^{(j)}}$ is either purely imaginary when $n$ is odd and $j$ is even, or is real otherwise. The resultant vector  $\RR$ of real components can then be defined as in equation (6) of the main text.

\section{SM8 -- Classification of eigenvalues of symmetric non-Hermitian Hamiltonians according to the sign of the discriminant.}
A classification of the eigenvalues of non-Hermitian Hamiltonians that satisfy a local anti-unitary symmetry given by Eq. (3) can be established according the sign of the discriminant $\Delta$ of their characteristic polynomial. 
This follows from the fact that polynomials of degree $n$ having real coefficients, and thus a real discriminant, satisfy the properties given in Table \ref{table:t2} for $n\geqslant 4$. The cases $n=2$ and $n=3$ are shown in table 1 of the main text.
A change of sign of the discriminant is thus associated with a change of the number of complex conjugate pairs of roots.
This result directly implies to the eigenvalues of PT-symmetric and pseudo-Hermitian Hamiltonians, since their characteristic polynomial has real coefficients, and the change of the number of complex conjugate pairs of eigenvalues is interprated as a spontaneously symmetry breaking.

For CP-symmetric and pseudo-chiral Hamiltonians, the coefficients $a_j$ of the characteristic polynomials satisfy $a_j=(-1)^{n+j}a_j^{*}$ (for $j=0,1\dots n$), and are thus not real in general. However, for a given $n$, those coefficient are either real or purely imaginary, depending on the parity of $j$. A change of variable $E\rightarrow -\ii E$ thus yields the polynomial equation $\tilde{P}(E)=0$ that has only real coefficients.

Let us detail the case $n=3$ with CP-symmetry as it is discussed in the main text. In that case, the even coefficients of the characteristic polynomial are imaginary and one can write $P(E) = a_3 E^3 + \ii \alpha_2 E^2 + a_1 E + \ii\alpha_0$ where $a_j=\ii \alpha_j$ for $j$ even, with $\alpha_j \in \mathbb{R}$. The change of variable $E\rightarrow -\ii E$ then yields $\tilde{P}(E)=0$ with $\tilde{P}(E) = a_3 E^3 - \alpha_2 E^2 - a_1 E + \ii\alpha_0$ which has real coefficients. Also, note that the discriminant is transformed as $\Delta \rightarrow -\Delta$ under this change of transformation. This justifies Table 1 for CP-symmetric and pseudo-chiral systems.

\begin{table}[h!]
\centering
\begin{tabular}{c|c}
 sgn $\Delta$         &  Pairing of roots \\
 \hline \hline
\multirow{3}*{$+1$}   & $2k$ pairs of complex conjugate roots  \\
                              &   $n-4k$ real roots     \\
                              & $k\leqslant n/4$ \\ 
\hline
$0$  & multiple roots \\
\hline
\multirow{3}*{$-1$}   & $2k-1$ pairs of complex conjugate roots  \\
                              &   $n-4k+2$ real roots     \\
                              & $k\leqslant (n-2)/4$ 
\end{tabular}
\caption{Pairing of roots for polynomials of degree $n\geqslant 4$ with real coefficients.}
\label{table:t2}
\end{table}

\section{SM9 -- More details on the shallow water model}

In the main text, we apply our theory on the concrete example of the  shallow water model in the presence of friction and rotation. 
\begin{figure}[h!]
\centering
\includegraphics[width=8.5cm]{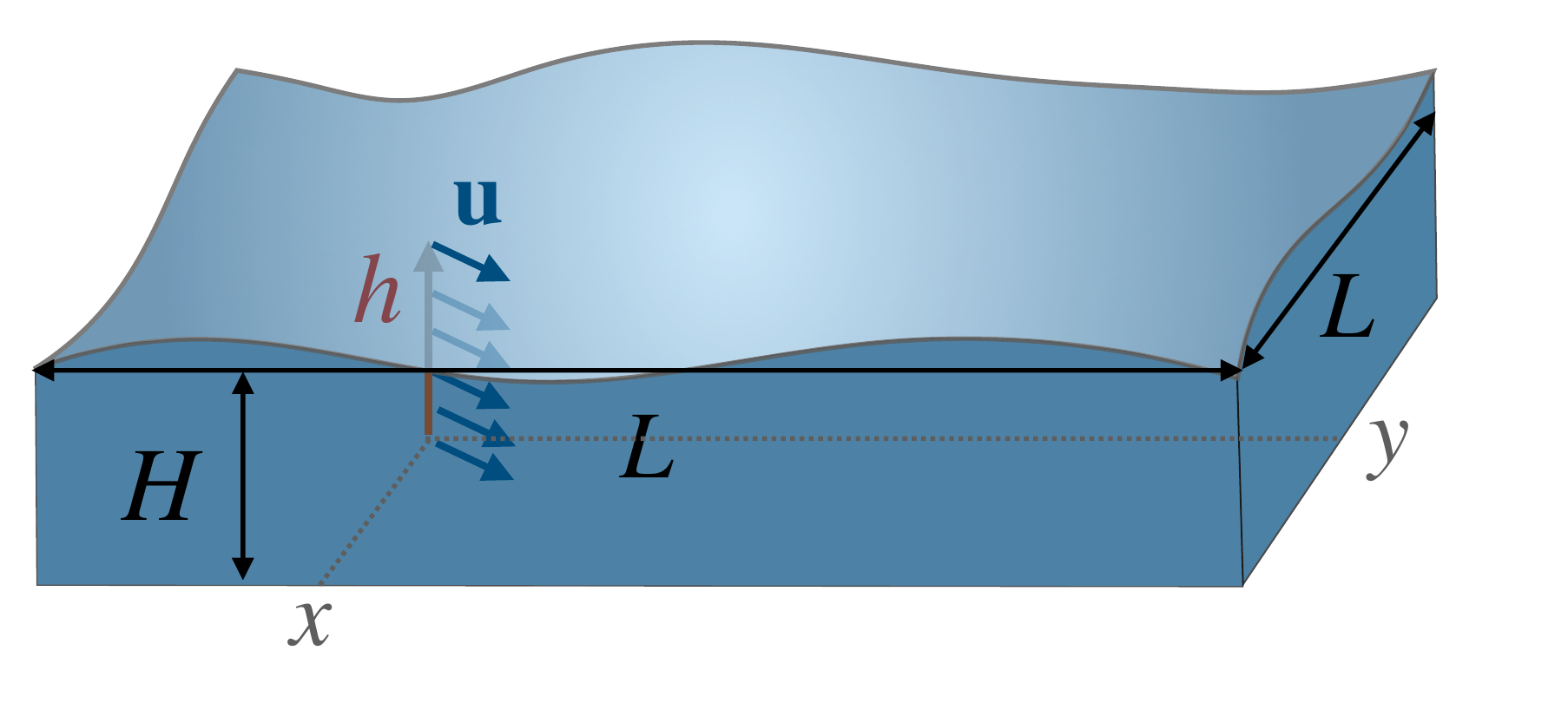}
\caption{ Sketch of the fluid described in the shallow water approximation. The thickness $H$ is assumed to be small compared to its length $L$, such that the fluid's velocity is constant in the vertical direction.}
\label{fig:SW}
\end{figure}

The rotating shallow water model describes the dynamics of a thin layer of an incompressible fluid. Here, ''thin' means that the vertical velocity of the fluid is neglected and that the horizontal velocity  is constant in the vertical direction, i.e. $\mathbf{u}(x,y,t) = (u_x(x,y,t) , u_y(x,y,t))$. It is for instance routinely used to describe atmospheric and oceanic waves over planetary distances, as the thickness of those fluids can be disregarded  when looking at the large scale dynamics  \cite{VallisBook}. Also, the fluid's thickness $h(x,y,t)$ is allowed to vary. 
The shallow water model then follows from the Euler equations 
\begin{align}
\partial_t h + \nabla . (h \mathbf{u}) = 0\\
\partial_t \mathbf{u} + (\mathbf{u}.\nabla)\mathbf{u} = -g \nabla h
\end{align}
 in that thin layer limit (with $g$ the standard gravity). 
 Owing to mass and momentum conservation, the velocity field $\mathbf{u}(x,y,t)$ couples to the variation of thickness $h(x,y,t)$ of the fluid layer. The shallow water model is thus a two dimensional model that involves three fields.
When the fluid is rotating, a Coriolis term $f$ that is proportional to the angular velocity and couples the two velocity components, must be added to the momentum conservation \cite{VallisBook}.

A linearization around a state of rest (zero velocity and constant thickness $H$, i.e. $\mathbf{u}\sim \delta \mathbf{u}$ and $h\sim H+\delta h$) leads to the Shr\"odinger-like Hermitian eigenvalue problem 
$H \Psi = E \Psi$ after a Fourier decomposition on the basis $\ee^{\ii (E t - k_x x -k_y y)}$, and where $\Psi$ is the three-vector of Fourier components of $(\delta u_x, \delta u_y, \delta h)$ with
\begin{equation}
H = 
    \begin{pmatrix}
   0 &   \ii f          &    k_x   \\
    -\ii f        &  0   &    k_y   \\
    k_x           &       k_y        &   0
    \end{pmatrix}  \ .
\end{equation}

In the main text, diagonal frictional terms, are added, to break Hermiticity while conserving CP symmetry.

\end{document}